# Strain tuning of electronic structure in $Bi_4Ti_3O_{12}$-$LaCoO_3$ epitaxial thin films

Woo Seok Choi[1,2,*] and Ho Nyung Lee[1,**]

[1]*Materials Science and Technology Division, Oak Ridge National Laboratory, Oak Ridge, TN 37831, USA*

[2]*Department of Physics, Sungkyunkwan University, Suwon 440-746, Korea*

[*]e-mail: choiws@skku.edu.

[**]e-mail: hnlee@ornl.gov.

We investigated the crystal and electronic structures of ferroelectric $Bi_4Ti_3O_{12}$ (BiT) single crystalline thin films site-specifically substituted with $LaCoO_3$ (LCO). The epitaxial films were grown by pulsed laser epitaxy on $NdGaO_3$ and $SrTiO_3$ substrates to vary the degree of strain. With increasing the LCO substitution, we observed a systematic increase in the *c*-axis lattice constant of the Aurivillius phase related with the modification of pseudo-orthorhombic unit cells. These compositional and structural changes resulted in a systematic decrease in the band gap, i.e., the optical transition energy between the oxygen 2*p* and transition metal 3*d* states, based on a spectroscopic ellipsometry study. In particular, the Co 3*d* state seems to largely overlap with the Ti $t_{2g}$ state, decreasing the band gap. Interestingly, the applied tensile strain facilitates the band gap narrowing, demonstrating that epitaxial strain is a useful tool to tune the electronic structure of ferroelectric transition metal oxides.

**I. INTRODUCTION**

Ferroelectric transition metal oxides (TMOs) with a low band gap ($E_g$) have recently attracted a lot of attention due to their promising opto-electronic and energy device applications [1-4]. Low $E_g$ ferroelectrics such as $BiFeO_3$ ($E_g$ = ~2.3 eV) [5] and hexagonal $R$MnO$_3$ ($R$ = rare earth, $E_g$ = ~1.5 eV) [6] exhibit intriguing physical properties, such as magnetoelectric coupling, switchable photovoltaic, and topological vortex domains, stemming from the spontaneous electric polarization [7-9]. We have recently introduced the site-specifically substituted $Bi_4Ti_3O_{12}$-$LaCoO_3$ (BiT-LCO) system as a new member of the low $E_g$ ferroelectrics [10,11]. In this system, unlike conventional TMOs, $E_g$ of the system could be largely reduced (~1 eV) by changing the LCO substitution rate [12]. $E_g$ could be decreased as low as ~2.55 eV, while maintaining the strong ferroelectricity of BiT [13]. Since $E_g$ determines most of the electronic and optical characteristics of a material, the ability to tune $E_g$ provides an unprecedented pathway to developing TMOs with novel properties.

The decrease of $E_g$ in the BiT-LCO system is found to be originating from the incorporation of Co ion [10,11]. Substitution of Bi with La greatly stabilized the ferroelectricity[14-17], but it does not have any influence on $E_g$ [18]. Specifically, density functional theory calculations indicate that the creation of an additional Co electronic state just below the conduction band of BiT decreases $E_g$ of the system [10,19]. The crucial role of Co has also been confirmed using bulk powder studies [20,21].

In this work, we impose a distinctive epitaxial strain to BiT-LCO thin films and investigate their crystal and electronic structures. As epitaxial strain is an efficient tuning parameter for various physical properties of TMOs [22-24], we were able to observe detailed crystal and electronic structure evolution with Co incorporation into BiT depending on the strain state. In particular, we observed a systematic change in the lattice parameters and a subsequent decrease in the optical transition energy between the electronic states with Co incorporation. Based on the experimental results, we discuss the relationship among the epitaxial strain, Co incorporation, lattice parameter, and electronic structure, which provide a deeper understanding on the $E_g$ reduction mechanism in BiT.

## II. EXPERIMENTAL

We used pulsed laser epitaxy to fabricate (001)-oriented BiT-LCO epitaxial films on single crystal orthorhombic (110) NdGaO$_3$ (NGO) and cubic (001) SrTiO$_3$ (STO) substrates. The two substrates were chosen as they provide both a superior platform for high quality thin film growth with excellent crystallinity and a reasonably small lattice mismatch (< 2%). The latter enables us to grow coherently strained films with different degrees of epitaxial strain. Samples were grown at 700 °C in 100 mTorr at the repetition rate of 10 Hz. The high repetition rate was used to prevent the loss of volatile Bi in the films during the growth [15,16,25]. We controlled the laser pulse number in order to systematically change the LCO concentration in the films. Quarter and half unit cell layers of LCO and quarter unit cell layers of BiT were alternatingly ablated from stoichiometric LCO and BiT targets to grow 1BiT-1LCO (1B1L) and 1BiT-2LCO (1B2L) alloys, respectively. 1B1L and 1B2L thin films have one and two pseudo-cubic unit cell layers of LCO incorporated into one pseudo-orthorhombic unit cell layer of BiT, respectively. More detailed growth method is described elsewhere [10]. The crystal structure of films was characterized using high resolution x-ray diffraction (XRD, X'Pert, Panalytical Inc.). Spectroscopic ellipsometry (M-2000, J. A. Woollam Co.) was used to obtain optical conductivity spectra ($\sigma_1(\omega)$) of the thin films between 1.5 and 5 eV at room temperature.

## III. RESULTS AND DISCUSSION

### A. Crystal structure of BiT-LCO epitaxial thin films

BiT belongs to the Aurivillius family, (Bi$_2$O$_2$)(Bi$_{n-1}$Ti$_n$O$_{3n+1}$) with $n$ = 3, of which the structure is built by alternating layers of one fluorite-like (Bi$_2$O$_2$)$^{2+}$ and two pseudo-perovskite (Bi$_2$Ti$_3$O$_{10}$)$^{2+}$ blocks. It has a highly anisotropic pseudo-orthorhombic unit cell ($a$ = 5.450, $b$ = 5.406, and $c$ = 32.832 Å) [26]. A pseudo-orthorhombic unit cell of BiT is schematically represented in Fig. 1(a). On the other hand, LCO is a simple perovskite with a rhombohedral unit cell ($a$ = 5.378 Å). The (in-plane) pseudo-cubic lattice parameters of the films and substrates are comparatively shown in the inset of Fig. 1(c). The (pseudo-)cubic lattice constants of NGO and STO substrates are 3.867 and 3.905 Å, respectively, which offer a good lattice match with BiT and LCO. Both NGO and STO substrates impose tensile strain to the film, with lattice mismatches of 0.7 and 1.7%, respectively. Although not presented in the current study, we note that imposing compressive strain would also provide an important aspect for strain-dependent optical property changes in the low band gap ferroelectrics.

Figures 1(b) and 1(c) show XRD patterns for BiT, 1B1L, and 1B2L films deposited on NGO and STO substrates, respectively. BiT peaks (002, 004, 006, etc.) due to the characteristic layered structure are shown with well-defined fringes, indicating homogeneous atomically flat surfaces and interfaces of the epitaxial thin films. The excellent crystalline quality let us calculate the total film thickness to be about 10 unit cells of BiT, by simply counting the number of fringes between the Bragg reflections. The x-ray rocking curve scans of the 006 reflection of 1B1L on NGO and STO substrates are shown in Figs. 1(d) and 1(e), respectively, confirming the excellent out-of-plane orientation of the films. Importantly, the peak structures in the XRD $\theta$-$2\theta$ scan are qualitatively the same between the pure BiT and BiT-LCO, indicating that the layered structure of BiT is robustly preserved upon alloying. All the peaks from the layered structure of BiT are clearly preserved for 1B1L and 1B2L, and those peaks were the only peaks present. The XRD result suggests that LCO is alloyed into BiT by substitution, instead of forming an additional layer or phase [10,11]. More specifically, based on the scanning transmission electron microscopy data, we can conclude that La substitutes for upper Bi in $Bi_2O_2$ layer, while Co substitutes for Ti near the $Bi_2O_2$ layer with the possible inclusion of oxygen vacancies, forming the self-ordered BiT-LCO superstructures of 1B1L and 1B2L, grown on both NGO and STO substrates [10]. We note that the generation of oxygen vacancies in Aurivillius phase materials is less influential for ferroelectricity [27]. From XRD $\theta$-$2\theta$ scans, most samples are confirmed to be epitaxially grown as a single crystal, except for the 1B2L film on NGO, which shows a small hint of foreign phases indicative of potential polycrystalline or impurity phases. However, the volume fraction of the impurity phases, if any, are negligible and do not contribute to the optical spectroscopy results.

The detailed structural evolution by alloying LCO into BiT is summarized in Fig. 2. In particular, the half-unit-cell lattice parameter, calculated from the 002 peak (Fig. 2(a)), systematically increases from ~16.37 to ~16.65 Å upon alloying LCO, for both substrates. Since the in-plane lattice parameter of LCO is smaller than BiT, the epitaxial tensile strain should decrease the *c*-axis lattice constant if a unit cell layer of LCO is epitaxially grown within the BiT matrix. The opposite trend suggests that the perovskite LCO is not forming a layer by itself. Instead, it indicates that La and Co substitutes for Bi and Ti elementally, forming an alloy of Bi-La-Ti-Co-O, with a self-ordered superstructure. We could also assume an increased number of oxygen vacancies with inclusion of $Co^{2+}$ or $Co^{3+}$ instead of $Ti^{4+}$, which could result in an expanded crystal lattice. Figure 2(b) shows the separation of the quarter-unit-cell lattice constants calculated from the separation of the 004 peak. Note that this separation of the 004 reflection was observed in the high quality pulsed laser epitaxy grown BiT thin films for the first time [10]. The separation of the peak for the BiT film suggests that even in the pure BiT, there is a modulation in the *c*/4 lattice constant, most likely due to the slight difference in the distance from upper and lower Bi to O within the $Bi_2O_2$ layer (*c'*/4 and *c"*/4 in Fig. 1(a)). With increasing LCO incorporation into BiT-LCO, the separation generally increases, possibly due to the disproportionate substitution of La for upper Bi within the $Bi_2O_2$ layer or inhomogeneous oxygen vacancy formation created to compensate the charge difference. We further note a quantitative difference in the quarter-unit-cell lattice modulation between the films grown on NGO and STO substrates, suggesting that the strain plays a role in determining the crystal structure. The STO substrate imposes larger tensile strain to the BiT film compared to the one on the NGO substrate, which might enhance the modulation in the *c*/4 lattice parameter.

Figure 2(c) shows full width at half maximum (FWHM) values of the 006 peak in rocking curve scans, as a function of the LCO incorporation. The FWHM values of the NGO and STO substrates are shown in lines for comparison. BiT-LCO thin films grown on NGO substrates have generally lower FWHM values compared to the ones on STO, owing to the better crystallinity of NGO substrates and lower lattice mismatch. On the other hand, the FWHM value considerably decreases with increasing incorporation of LCO, despite the induced disorder in the crystal. This result suggests that the *c*-axis orientation of the film is actually better for the BiT-LCO films than the pure BiT films on both substrates.

**B. Electronic structure and band gap evolution**

The electronic structure of BiT-LCO thin films including $E_g$ has been identified using spectroscopic ellipsometry. Figure 3 shows the real part of the optical conductivity spectra as a function of photon energy ($\sigma_1(\omega)$). We could further fit $\sigma_1(\omega)$ using Lorentz oscillators,

$$\sigma_1(\omega) = \frac{e^2}{m^*} \sum_j \frac{n_j \gamma_j \omega^2}{\left(\omega_j^2 - \omega^2\right)^2 + \gamma_j^2 \omega^2},$$

where $e$, $m^*$, $n_j$, $\gamma_j$ and $\omega_j$ are the electronic charge, the effective mass, the carrier density, the scattering rate, and the resonant frequency of the *j*-th oscillator, respectively. Light blue solid line is the example of the fit result shown for 1B2L films grown on NGO and STO substrates, indicating a good fitting quality. The values of the fitting parameters are listed in Table I.

Undoped BiT has only $Ti^{4+}$ ($3d^0$) ions, which construct unoccupied $t_{2g}$ and $e_g$ states based on the octahedral crystal field. The two Lorentz oscillators at ~3.8 (*α*) and ~4.8 eV (*β*) can be attributed to the charge-transfer transition between O 2*p* and Ti 3*d* $t_{2g}$ and $e_g$ states, respectively. These two transitions are sufficient to construct $\sigma_1(\omega)$ for the undoped BiT films on both NGO and STO substrates. Note that the transition from O 2*p* state to Ti 3*d* $t_{2g}$ state defines the $E_g$ (denoted as arrows on the *x*-axis) of ~3.5 eV for both BiT films [28,29].

By site-specific substitution with $LaCoO_3$, it has been suggested that additional Co state plays an important role in decreasing $E_g$. It has been clarified in different systems (single-crystalline film and poly-crystalline powders) that Co is indeed responsible for the change in the electronic structure, and La does not play any role in changing $E_g$ [10,11,20,21]. Unoccupied Co state is expected to be formed just above the Fermi level ($E_F$) and below Ti 3*d* $t_{2g}$ state, which decreases $E_g$ [10]. However, the detailed electronic structure evolution with the incorporation of Co was rather elusive.

We note that better identification of the electronic structure evolution was possible by growing BiT-LCO thin films on NGO. In particular, as shown in Fig. 3(a), the optical transition peaks can be much better resolved for the films grown on NGO, than those grown on STO. Indeed, the average values of $\gamma_\alpha$ and $\gamma_\beta$ are ~0.4 and ~1.2 eV, respectively, for the films grown on NGO, while the values are ~0.7 and ~2.1 eV, respectively, for the films grown on STO. Note that $\gamma$ represents the FWHM of each peak in $\sigma_1(\omega)$. The sharpness of the peaks indicates well-defined electronic structure, which might originate from the excellent crystalline quality of the films grown on NGO substrates. We also note the absence of optical transitions from Co impurity phases, such as CoO or $Co_3O_4$, indicating our optical spectroscopic results are free from the contribution from possible impurity phases, if any.

With the incorporation of Co, a new peak (optical transition between O 2$p$ and the emergent Co 3$d$ state) below $\alpha$ is expected in $\sigma_1(\omega)$. However, as shown in Fig. 3, the results indicate that only two Lorentz oscillators are sufficient to fit $\sigma_1(\omega)$ of all the BiT-LCO thin films. While it is evident that peak $\alpha$ can be deconvoluted into as many Lorentz oscillators as possible, it would only give rise to an arbitrary conclusion. Nevertheless, we can conclude that the emergent occupied Co 3$d$ state is very closely located to the Ti 3$d$ $t_{2g}$ state, so that the two optical transitions cannot be distinguished. Another possible explanation of the two peak structure in $\sigma_1(\omega)$ is that Ti 3$d$ $t_{2g}$ state shifting to lower energy with the incorporation of Co 3$d$ state. Importantly, while the peak $\alpha$ shifts to a lower photon energy value, the peak $\beta$ stays almost at the same energy as a function of Co incorporation. Within the conventional octahedral crystal field, it is not likely that only the $t_{2g}$ state shifts to a lower energy value while the $e_g$ state stays at the same photon energy. However, a similar orbital selective shift in the electronic structure has been reported for low dimensional titanate heterostructures [30]. Further studies regarding the orbital symmetry of the 3$d$ electrons of both Ti and Co might elucidate the intriguing phenomena.

In Fig. 4, we plot important parameters obtained from Lorentz oscillator fitting $\sigma_1(\omega)$. Figure 4(a) shows the peak positions ($\omega_j$) of each peak and $E_g$. As discussed above, $\omega_\alpha$ shows a drastic red-shift with increasing Co incorporation, while $\omega_\beta$ does not change as much. The decrease of the $\omega_\alpha$ defines the decreasing $E_g$. The $E_g$ values are 3.4 (3.55), 3.25 (3.3), and 2.9 (2.65) eV for BiT, 1B1L and 1B2L grown on NGO (STO) substrates, respectively. While the decrease in $\omega_\alpha$ and $E_g$ with Co incorporation is universal, we note that the films on NGO substrates show a more subtle change compared to the films on STO substrates. This result should be related to the larger structural modification in BiT-LCO films on STO with introducing LCO, compared to the films on NGO, and certainly indicates the substantial role of epitaxial strain in determining the electronic structure.

A spectral weight ($SW \equiv \int \sigma_1(\omega)\, d\omega$) analysis shown in Fig. 4(b) provides more insight into intriguing epitaxial strain-dependent electronic structure evolution. For the BiT-LCO films grown on NGO, $SW_\alpha$ increases while $SW_\beta$ decreases with introducing Co. As peak $\alpha$ is partially attributed to the transition from the increasing Co 3$d$ states, this is a systematic trend. However, the BiT-LCO films grown on STO show exactly the opposite trend. Here, $SW_\alpha$ decreases with the introduction of Co whereas $SW_\beta$ increases. We note that $SW$ analyses for the films grown on STO might be rather inaccurate, especially due to the large $\gamma_j$ values. For example, there is a possibility that peak $\beta$ of the 1B2L film on STO might further be deconvoluted to result in a total of three optical transitions. However, $\sigma_1(\omega)$ of the BiT-LCO films on NGO cannot be understood with the same scenario, as the peak $\alpha$ is clearly separated from peak $\beta$ for all the films. In order to understand the anomalous $SW$ evolution, we suggest a moderate mixture of $t_{2g}$ and $e_g$ states for the films grown on STO. Indeed, the larger tensile strain and larger separation of $c$/4 lattice peaks for the films grown on STO (Fig. 2(b)) indicate a larger lattice distortion of the perovskite blocks. This might lead to a substantial change in the octahedral crystal field and even might partially reverse the energy levels of $t_{2g}$ and $e_g$ states. Moreover, the increased number of oxygen vacancies with increasing Co substitution might even result in a tetrahedral, instead of octahedral, crystal field. Since the $e_g$ states are lower than the $t_{2g}$ states in energy for the tetrahedral crystal field, the mixture of $t_{2g}$ and $e_g$ states might unexpectedly be realized.

## VI. SUMMARY

In summary, we have investigated the structural and optical properties of $LaCoO_3$-substituted ferroelectric $Bi_4Ti_3O_{12}$ grown epitaxially on $NdGaO_3$ and $SrTiO_3$ substrates. The electronic structure evolution, which yields a reduced band gap, is closely related to a complex balance among the crystal structure, epitaxial strain, and film quality. Our study suggests that a careful design of heterostructures composed of correlated oxide ferroelectrics can lead to a novel materials system, in which the band gap tunability help develop oxide heterostructures with technologically more desirable functionalities.

**ACKNOWLEDGEMENTS**

This work was supported by the U.S. Department of Energy, Office of Science, Basic Energy Sciences, Materials Sciences and Engineering Division (sample synthesis and characterization). W.S.C. was supported by Basic Science Research Program through the National Research Foundation of Korea (NRF) funded by the Ministry of Science, ICT and future Planning (NRF-2014R1A2A2A01006478) (structural and optical data analyses).

Figure Captions

FIG. 1. (Color online) Crystal structure of single crystal BiT-LCO thin films. (a) Schematic diagram of a pseudo-orthorhombic unit cell of BiT. X-ray diffraction $\theta$-$2\theta$ patterns of BiT (red), 1B1L (green), and 1B2L (blue) epitaxial thin films on (b) NGO and (c) STO substrates reveal very strong reflections with well-defined fringes indicating the excellent film quality. The substrate peaks are indicated as [*]. The Aurivillius structure represented in (a) is clearly observed for all the films. The inset in (b) shows the pseudo-cubic lattice constant values of the substrates and the constituent compounds of the thin film. Narrow peaks from x-ray rocking curve scans of the 006 reflection of 1B1L on (d) NGO and (e) STO substrates confirm the excellent crystal quality.

FIG. 2. (Color online) Structural parameters of BiT-LCO thin films. (a) Unit cell parameter corresponding to $c/2$ of BiT, (b) difference between $c'/4$ and $c''/4$ (calculated from the separation of the 004 peak), and (c) FWHM from the rocking curve peak of the 006 reflection of BiT-LCO epitaxial thin films, as a function of the LCO substitution. As more LCO is incorporated into the film, the lattice expands along the $c$-axis, and the difference between $c'/4$ and $c''/4$ increases. The LCO substitution yields a better $c$-axis orientation and crystalline quality, as seen from the decreased FWHM with $x$.

FIG. 3. (Color online) $\sigma_1(\omega)$ and electronic structure. $\sigma_1(\omega)$ of BiT (red), 1B1L (green), and 1B2L (blue) epitaxial thin films on (a) NGO and (b) STO substrates. All the $\sigma_1(\omega)$ could be deconvoluted using two Lorentz oscillators denoted as $\alpha$ (dashed black line) and $\beta$ (thin black line). As an example, resultant fitting curves for 1B2L films on both NGO and STO substrates are shown as light blue lines. The arrows at the upper $x$-axis indicate the optical band gap of the films.

FIG. 4. (Color online) Evolution of optical spectroscopic characteristic and electronic structure of BiT-LCO thin films. (a) $\omega_j$ and (b) $SW$ of the Lorentz oscillators as a function of the LCO substitution.

A clear and consistent peak shift with the incorporation of LCO can be observed for the films grown on NGO and STO, along with the decreased $E_g$. *SW*, however, shows an anomalous behavior, which is the opposite for the films grown on NGO and STO substrates.

TABLE I. Summary of Lorentzian parameters used for fitting $\sigma_1(\omega)$ of BiT-LCO thin films.

|        |      | $\omega_\alpha$ | $SW_\alpha$ | $\gamma_\alpha$ | $\omega_\beta$ | $SW_\beta$ | $\gamma_\beta$ |
|--------|------|------|------|------|------|-------|------|
| on NGO | BiT  | 3.67 | 4.44 | 0.25 | 4.87 | 60.48 | 1.25 |
|        | 1B1L | 3.51 | 5.79 | 0.40 | 4.80 | 52.99 | 1.20 |
|        | 1B2L | 3.43 | 7.65 | 0.55 | 4.75 | 45.13 | 1.20 |
| on STO | BiT  | 3.95 | 8.58 | 0.70 | 4.87 | 74.71 | 1.50 |
|        | 1B1L | 3.55 | 3.78 | 0.80 | 4.84 | 83.86 | 2.00 |
|        | 1B2L | 3.35 | 2.24 | 0.65 | 4.7  | 86.15 | 2.8  |

References

[1]     H. Huang, Nat. Photon. **4**, 134 (2010).

[2]     I. Grinberg, D. Vincent West, Maria Torres, Gaoyang Gou, David M. Stein, Liyan Wu, Guannan Chen, Eric M. Gallo, Andrew R. Akbashev, Peter K. Davies, Jonathan E. Spanier, and Andrew M. Rappe, Nature **503**, 509 (2013).

[3]     J. Seidel and L. M. Eng, Curr. Appl. Phys. **14**, 1083 (2014).

[4]     R. Nechache, C. Harnagea, S. Li, L. Cardenas, W. Huang, J. Chakrabartty, and F. Rosei, Nat. Photon. **9**, 61 (2015).

[5]     J. F. Ihlefeld *et al.*, Appl. Phys. Lett. **92**, 142908 (2008).

[6]     W. S. Choi *et al.*, Phys. Rev. B **77**, 045137 (2008).

[7]     T. Choi, S. Lee, Y. J. Choi, V. Kiryukhin, and S.-W. Cheong, Science **324**, 63 (2009).

[8]     S. Y. Yang, J. Seidel, S. J. Byrnes, P. Shafer, C.-H. Yang, M. D. Rossell, P. Yu, Y.-H. Chu, J. F. Scott, J. W. Ager, III, L. W. Martin, and & R. Ramesh, Nat. Nano. **5**, 143 (2010).

[9]     T. Choi, Y. Horibe, H. T. Yi, Y. J. Choi, W. Wu, and S. W. Cheong, Nat. Mater. **9**, 253 (2010).

[10]    W. S. Choi, M. F. Chisholm, D. J. Singh, T. Choi, G. E. Jellison, and H. N. Lee, Nat. Commun. **3**, 689 (2012).

[11]    W. S. Choi and H. N. Lee, Appl. Phys. Lett. **100**, 132903 (2012).

[12]    T. Arima, Y. Tokura, and J. B. Torrance, Phys. Rev. B **48**, 17006 (1993).

[13]    S. E. Cummins and L. E. Cross, Appl. Phys. Lett. **10**, 14 (1967).

[14]    B. H. Park, B. S. Kang, S. D. Bu, T. W. Noh, J. Lee, and W. Jo, Nature **401**, 682 (1999).

[15]    H. N. Lee, D. Hesse, N. Zakharov, and U. Gösele, Science **296**, 2006 (2002).

[16]    H. N. Lee, D. Hesse, N. Zakharov, S. K. Lee, and U. Gösele, J. Appl. Phys. **93**, 5592 (2003).

[17]    Y. Shimakawa, Y. Kubo, Y. Tauchi, H. Asano, T. Kamiyama, F. Izumi, and Z. Hiroi, Appl. Phys. Lett. **79**, 2791 (2001).

[18]    Z. G. Hu, Y. W. Li, F. Y. Yue, Z. Q. Zhu, and J. H. Chu, Appl. Phys. Lett. **91**, 221903 (2007).

[19]    D. J. Singh, S. S. A. Seo, and H. N. Lee, Phys. Rev. B **82**, 180103 (2010).

[20]    C. Bark, Met. Mater. Int. **19**, 1361 (2013).

[21]    C. W. Bark, J. Nanoelectron. Optoelectron. **8**, 454 (2013).

[22]    D. G. Schlom, L.-Q. Chen, C. J. Fennie, V. Gopalan, D. A. Muller, X. Pan, R. Ramesh, and R. Uecker, MRS Bull. **39**, 118 (2014).

[23]    W. S. Choi, J.-H. Kwon, H. Jeen, J. E. Hamann-Borrero, A. Radi, S. Macke, R. Sutarto, F. He, G. A. Sawatzky, V. Hinkov, M. Kim, and H. N. Lee, Nano Lett. **12**, 4966 (2012).

[24]    J. Chakhalian, J. M. Rondinelli, J. Liu, B. A. Gray, M. Kareev, E. J. Moon, N. Prasai, J. L. Cohn, M. Varela, I. C. Tung, M. J. Bedzyk, S. G. Altendorf, F. Strigari, B. Dabrowski, L. H. Tjeng, P. J. Ryan, and J. W. Freeland, Phys. Rev. Lett. **107**, 116805 (2011).

[25]    H. N. Lee, S. Senz, A. Visinoiu, A. Pignolet, D. Hesse, and U. Gösele, Appl. Phys. A **71**, 101 (2000).

[26] A. D. Rae, J. G. Thompson, R. L. Withers, and A. C. Willis, Acta Crystallogr. Sect. A **46**, 474 (1990).

[27] J. F. Scott, *Ferroelectric Memories* (Springer, Heidelberg, Germany, 2000).

[28] S. Ehara, K. Muramatsu, M. Shimazu, J. Tanaka, M. Tsukioka, Y. Mori, T. Hattori, and H. Tamura, Jpn. J. Appl. Phys. **20**, 877 (1981).

[29] C. Jia, Y. Chen, and W. F. Zhang, J. Appl. Phys. **105**, 113108 (2009).

[30] S. S. A. Seo, M. J. Han, G. W. J. Hassink, W. S. Choi, S. J. Moon, J. S. Kim, T. Susaki, Y. S. Lee, J. Yu, C. Bernhard, H. Y. Hwang, G. Rijnders, D. H. A. Blank, B. Keimer, and T. W. Noh, Phys. Rev. Lett. **104**, 036401 (2010).

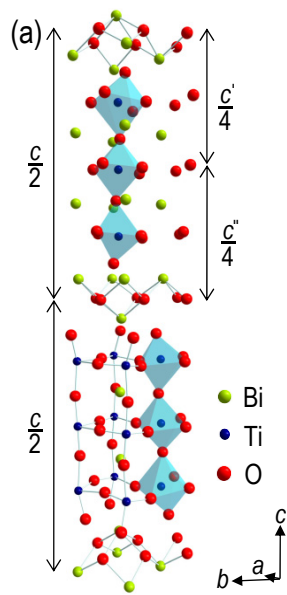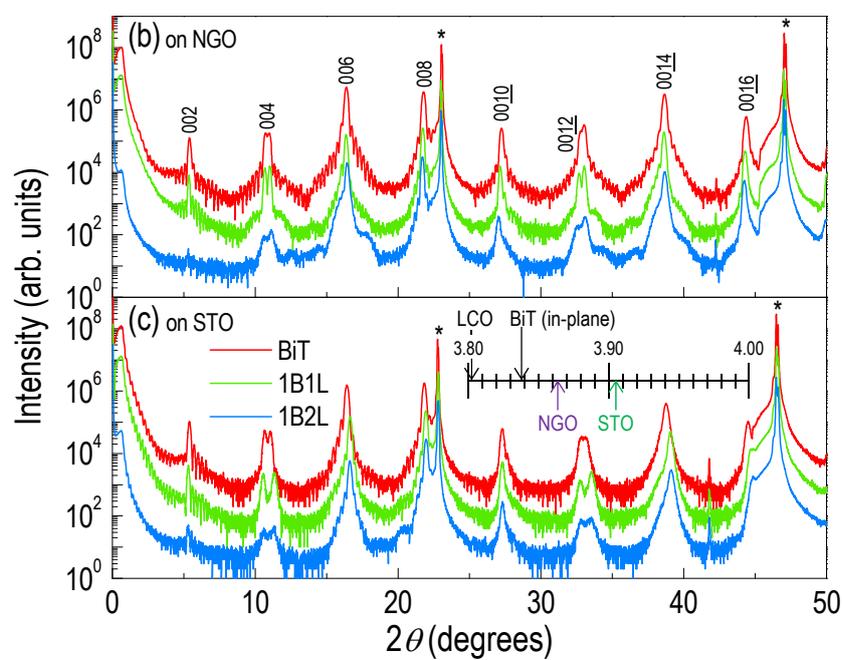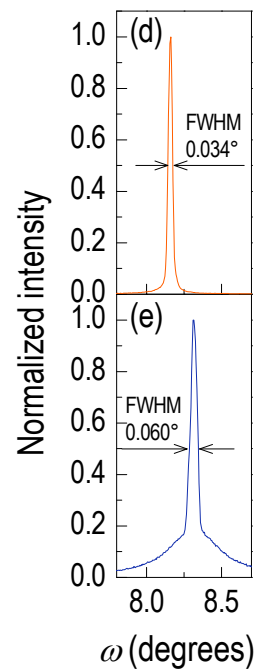

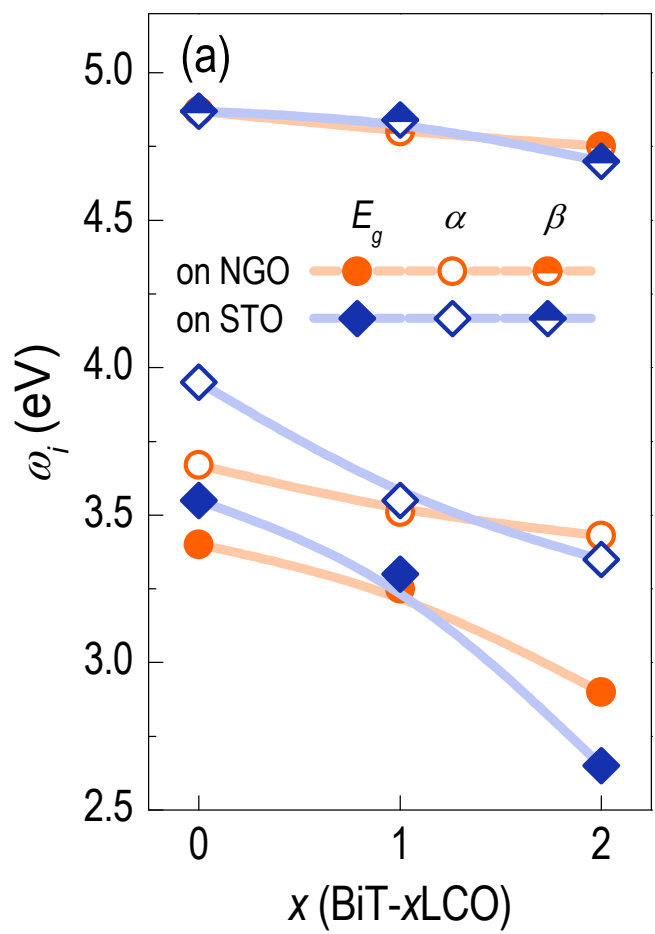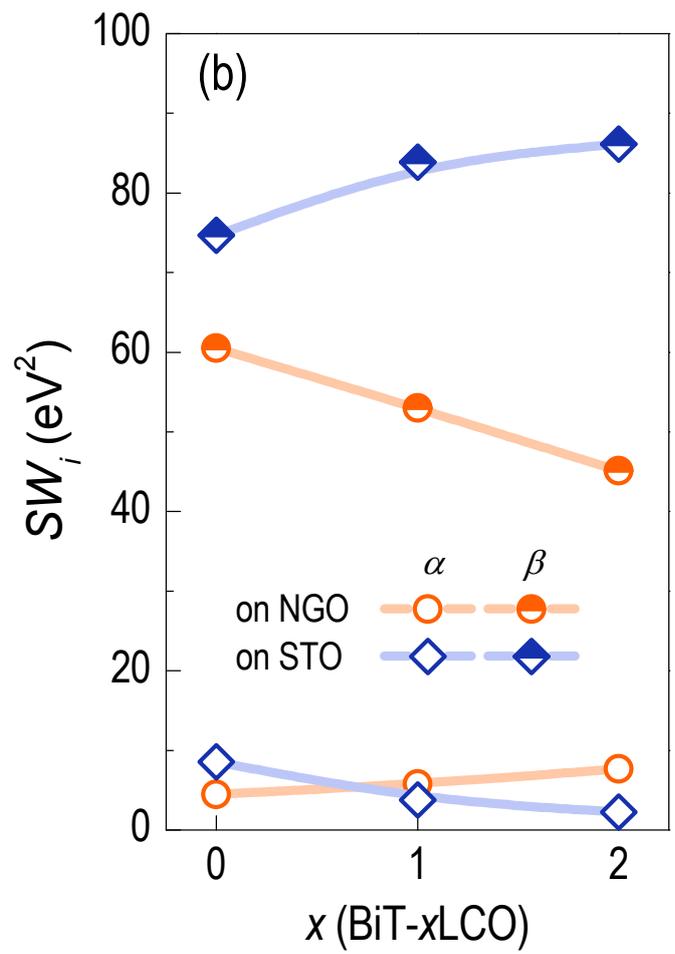

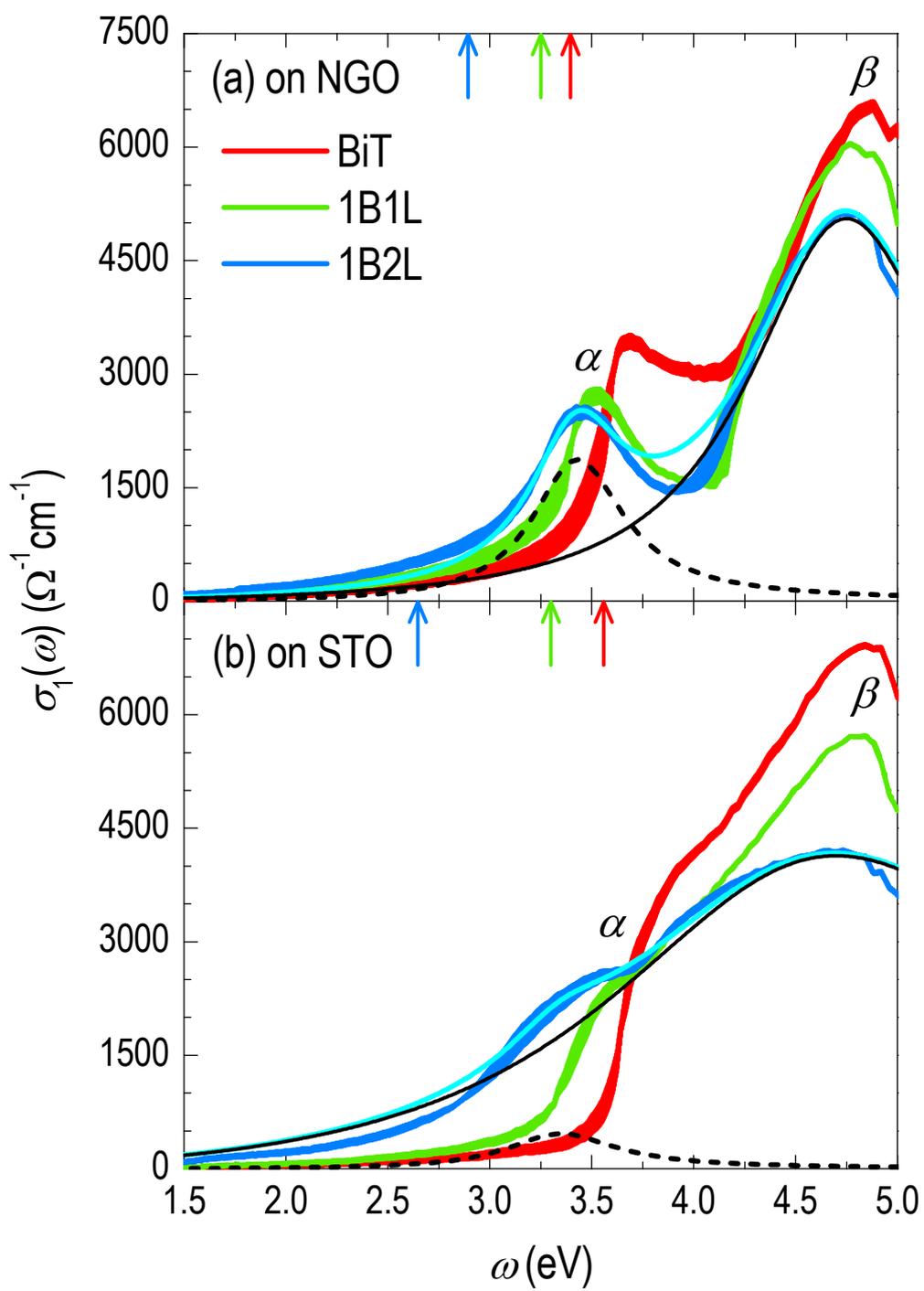

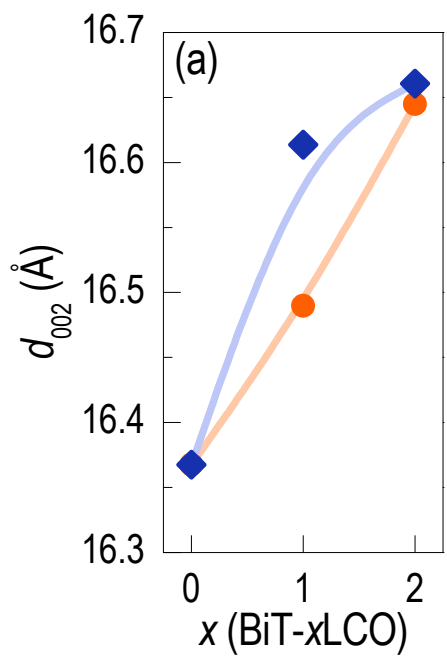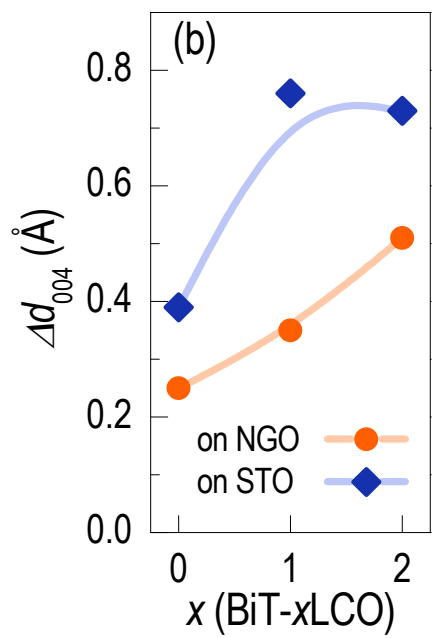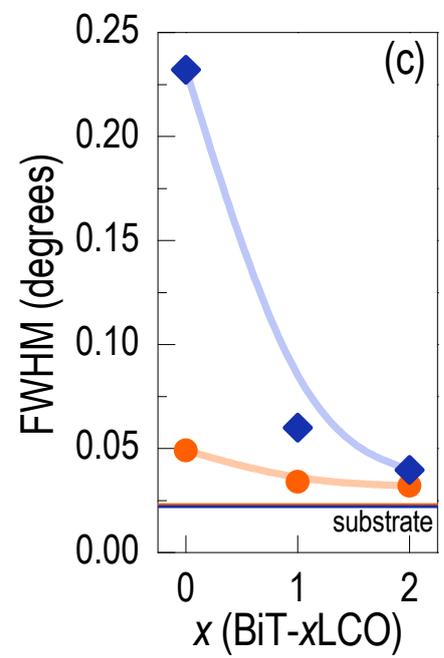